\documentclass[prl,twocolumn,amsmath,amssymb]{revtex4}
\usepackage{graphicx}% Include figure files
\usepackage{dcolumn}% Align table columns on decimal point
\usepackage{bm}% bold math
\begin{document}

%\draft

%\twocolumn[\hsize\textwidth\columnwidth\hsize\csname@twocolumnfalse\endcsname

\title{Universal dynamical conductance in graphite}

\author{A. B. Kuzmenko, E. van Heumen, F. Carbone\footnote{Present address:
California Institute of Technology, Pasadena, CA 91125, USA}
 and D. van der Marel}

\affiliation{DPMC, University of Geneva, 1211 Geneva 4,
Switzerland\\}

\begin{abstract}

We find experimentally that the optical sheet conductance of graphite per graphene layer is very close to $(\pi/2)e^2/h$, 
which is the theoretically expected value of dynamical conductance of isolated monolayer graphene.
Our calculations within the Slonczewski-McClure-Weiss model explain well why the interplane hopping
leaves the conductance of graphene sheets in graphite almost unchanged for photon energies between 0.1 and 0.6 eV, even though it significantly
affects the band structure on the same energy scale. The f-sum rule analysis shows that the large increase of the Drude spectral weight
as a function of temperature is at the expense of the removed low-energy optical spectral weight of transitions between hole and electron bands.

\end{abstract}

\maketitle

\begin{figure}[htb]
   \centerline{\includegraphics[width=8cm,clip=true]{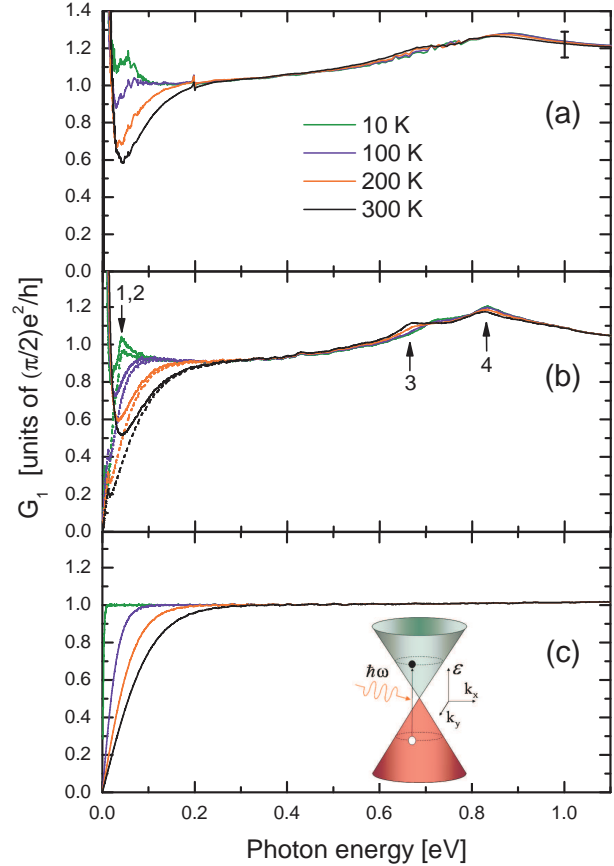}}
   \caption{The real part of the optical sheet conductance of graphite per layer (a - experiment, b - calculation) as well as the calculated
   conductance of isolated undoped graphene (c). The error bar indicates the inaccuracy of the absolute value due to systematic
   experimental uncertainties; the relative inaccuracy of the temperature dependence is much
   smaller. In panel (b) the dashed lines indicate the interband conductance only, while the solid
   line represent the total conductance with a Drude peak added (the scattering rate is 5 meV). In panel (c) the thermally activated Drude peak is
   not shown. The arrows in panel (b) correspond to optical transitions indicated in figure \ref{FigBands}. A certain noise in panels
   (b) and (c) is of numerical origin. The inset of panel (c) depicts the optical transitions between hole and electron bands in monolayer graphene.}
   \label{FigG1}
\end{figure}

One of the most remarkable macroscopic manifestations
of quantum mechanics is the appearance of a universal
conductance $e^2/h$, where $e$ is the elementary charge and $h$
is the Planck constant, in various physical phenomena. This
value appears in the quantum Hall effect
\cite{vonKlitzingPRL80,LaughlinPRB81}, in the
superconductor-insulator transition in two dimensions
\cite{JaegerPRB86,FisherPRL90} and in 1D ballistic transport
\cite{ButtikerPRB85,vanWeesPRL88,WharamJPC88}. Notably, all of these
observations were restricted so far to the DC transport. Monolayer graphene \cite{NovoselovScience04,GeimNM07}
represents an interesting example, where the optical, or AC, conductance due to optical interband transitions
is expected to be frequency independent and solely determined by the same universal value
\cite{AndoJPSJ02,GusyninPRL06,FalkovskyEPJB07} in a broad range of photon energies:
\begin{equation}\label{G0}
G_{1}(\omega)=G_{0}\equiv\frac{e^2}{4\hbar} \approx
6.08\cdot10^{-5} \ \Omega^{-1}
\end{equation}
\noindent (index '1' refers to the real part). 
Quite remarkably, $G_{1}(\omega)$ does not depend on
microscopic parameters that normally determine optical
properties of materials. This is a consequence of the unusual low-energy
electronic structure that resembles the dispersion of
relativistic particles \cite{GeimNM07}. At energies
considerably smaller than the bandwidth ($\lesssim$ 2 eV), the
dispersion of monolayer graphene features massless
electron and hole conical bands $\epsilon_{e,h}({\bf k})=\pm
\hbar v_{F}|{\bf k}-{\bf k}_{D}|$ formed by the $p_{z}$
orbitals (as shown in the inset of Fig.\ref{FigG1}c), where ${\bf k}_{D}$ is the momentum of the Dirac point
(there are two of them at the points K and K' of the Brillouin zone) and $v_{F}\approx 10^6$ m/s is the Fermi velocity.
This type of dispersion is qualitatively different from more common quadratic massive bands, as has been
most convincingly demonstrated by a $E_{n}(B)\propto\mbox{sign}(n)\sqrt{|n|B}$
field dependence of Landau levels
\cite{NovoselovNature05,ZhangNature05,SadowskiPRL06,JiangPRL07}.

The absolute value of the optical conductance in graphene,
the determination of which is an experimentally challenging task, has not yet been reported.
However, it is legitimate to ask whether the predicted universality can already be observed
in the conventional bulk graphite. Here we experimentally show that the answer is
affirmative and explain it using the classical Slonczewski-Weiss-McClure (SWMcC) band model of
graphite \cite{SWMcC}.

It is instructive to begin with a short summary of expected optical properties
of monolayer graphene. Eq.(\ref{G0}), apart from the numerical factor, follows from a simple
dimensional analysis. When the chemical
potential $\mu$ is zero, the conductance is given by the formula:
\begin{equation}\label{KG}
G_{1}(\omega)=\frac{\pi e^2}{\omega}|{\bf v}(\omega)|^2
D(\omega)\left[f\left(-\frac{\hbar\omega}{2}\right)-f\left(\frac{\hbar\omega}{2}\right)\right]
\end{equation}
\noindent where ${\bf v}(\omega)$ is the velocity matrix
element between the initial state with energy
-$\hbar\omega/2$ and the final state with energy
$\hbar\omega/2$, $D(\omega)$ is the 2D joint density
of states and $f(\epsilon)=\left[\exp(\epsilon/T)+1\right]^{-1}$ is the Fermi-Dirac distribution.
If only the nearest neighbor hopping is present
then $D_{i,f}(\omega)\propto \hbar\omega/t^{2}a^{2}$ and
$|{\bf v}_{i,f}(\omega)|\propto v_{F}\propto ta/\hbar$, where $a\approx$ 1.42 \AA\ is
the interatomic distance and $t\approx$ 3 eV is the hopping value.
Therefore the non-universal parameters $t$, $a$ as well as the frequency $\omega$ in
equation (\ref{KG}) cancel and one obtains at zero temperature $G_{1}(\omega)\propto e^2/\hbar$.
An interesting consequence of equation (\ref{G0}) is that the
optical transmittance of a free standing monolayer graphene
sample is also frequency independent and is expressed solely
via the fine structure constant $\alpha=e^2/\hbar c$:
\begin{equation}
 T_{opt}=\left(1+\frac{\pi\alpha}{2}\right)^{-2}\approx 1-\pi\alpha\approx
 0.977
\end{equation}
as follows from the Fresnel equations in the thin-film limit.
The calculated conductance of undoped graphene is shown in
figure \ref{FigG1}c for several temperatures \cite{RemarkCalculation}. The depletion of
the low-energy conductance with temperature is due to the
gradual equilibration of the electron and hole occupation
numbers close to the Fermi level. The 'removed' optical
spectral weight accumulates at zero frequency as a Drude peak (not shown),
whose integrated intensity increases linearly with temperature
\cite{PedersenPRB03,FalkovskyEPJB07}:
\begin{equation}\label{Dgraphene}
D_{\mbox{\scriptsize graphene}}(T)=\frac{e^2}{\hbar}T\ln 2
\end{equation}

\begin{figure}[htb]
   \centerline{\includegraphics[width=8.5cm,clip=true]{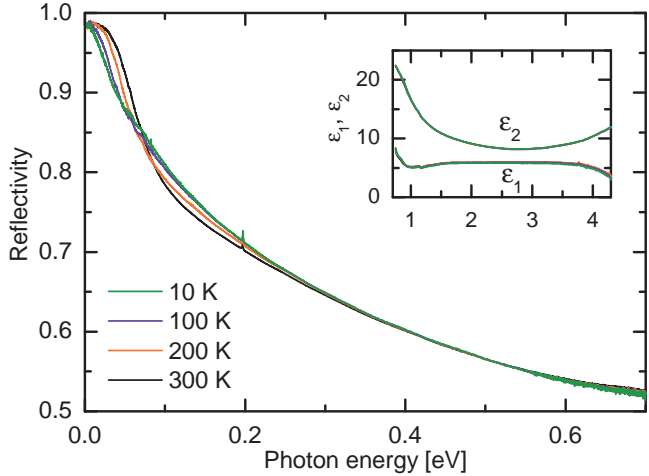}}
   \caption{In-plane infrared reflectivity spectra of highly ordered pyrolytic graphite.
   The inset shows the real and the imaginary parts of the dielectric function obtained by ellipsometry at higher energies.}
   \label{FigRefl}
\end{figure}

Optical measurements from 10 to 300 K were performed on highly
ordered pyrolytic graphite (HOPG) of ZYA grade with a c-axis
misorientation smaller than 0.4$^{\circ}$. Samples were cleaved
right before being inserted into a cryostat. Reflectivity at
near-normal incidence $R(\omega)$ was measured from 2 meV to
0.8 eV using a Fourier transform spectrometer (Fig. \ref{FigRefl}). The absolute
value was obtained by {\em in-situ} gold evaporation. Our
reflectivity spectra are in agreement with previous measurements
\cite{TaftPR65,GuizzettiPRL73,LiPRB06}. In the range
0.7 - 4.5 eV, we extracted the real and the imaginary parts of
the dielectric function from ellipsometric measurements (shown in the inset).
The correction for the admixture of the interplane optical
component to the ellipsometry spectra was initially performed using the
c-axis dielectric function found in Ref.
\cite{GreenawayPR69} and refined by comparing the reflectivity and ellipsometric spectra
in the range where they overlap. The uncertainty due to this procedure is the main source
of the error bar shown in figure \ref{FigG1}a. The complex in-plane conductivity $\sigma(\omega)$ in the whole
range was derived using a Kramers-Kronig consistent procedure \cite{KuzmenkoRSI05}, where the phase of the complex
reflectivity at low energies is anchored by ellipsometric data.
The sheet conductance per graphene layer was calculated using
the relation $G(\omega) = d_{c}\sigma(\omega)$, where $d_{c}$ =
3.35 \AA\ is the interlayer distance. Importantly, the optical
measurements reflect mostly bulk material properties, since the penetration
depth is several tens of $d_{c}$.

Figure \ref{FigG1}a shows the real part of the measured
conductance of HOPG normalized by $G_{0}$. One can notice 
a remarkable similarity between these
spectra and the calculated ones for graphene (panel c). The
conductance is almost constant and close to $G_{0}$, especially
between 0.1 and 0.6 eV. The second observation is that the
conductance at low energies shows a strong depletion with increasing
temperature in a fashion very similar to the temperature
dependence of $G_{1}(\omega)$ in graphene. In contrast to
the graphene spectra, the conductance of graphite shows a Drude
peak below 10-20 meV, an extra structure at about 50 meV and
two broad peaks at about 0.7 and 0.9 eV. The small narrow peak
at 0.2 eV is an optical phonon \cite{BrillsonJPCS71}. In
general, we conclude that the universality of the conductance
envisaged for the isolated graphene is also present graphite in a broad
energy range, in spite of the modification of the band structure
by a significant c-axis hopping ($t_{\perp} \approx$ 0.3 eV) \cite{SWMcC}. 
This result is not trivial since the universal conductance is observed at energies
of the order of $t_{\perp}$. For example, the calculated sheet
conductance of bilayer graphene shows a strong frequency
dependence caused by the interlayer hopping
\cite{NilssonPRL06,AbergelPRB07}.

\begin{figure}[htb]
   \centerline{\includegraphics[width=8cm,clip=true]{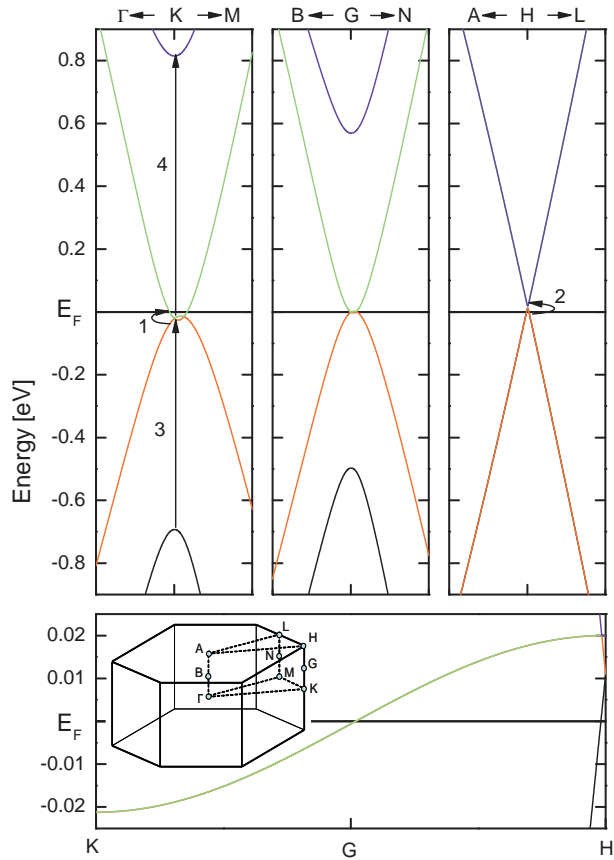}}
   \caption{The calculated dispersion of $p_{z}$ bands in graphite along the K-H line (along the K'-H' line the dispersion is the same).
   Different bands are shown with different colors. The arrows indicate interband transitions which form the correspondingly marked 
   peaks in figure \ref{FigG1}b.}
   \label{FigBands}
\end{figure}

In order to understand this observation, we calculated infrared spectra
of graphite with the standard AB stacking using a tight-binding Hamiltonian, which contains all hopping terms
of the SWMcC model \cite{SWMcC} and the
on-site energy difference between the non-equivalent carbon
atoms. The particular values of parameters, which were assumed to be temperature independent,
were taken from Ref. \cite{PartoensPRB06}.
The optical conductivity due to the direct interband
transitions and the Drude (intraband) spectral weight were obtained using the standard
relations, taking the temperature dependent occupation numbers
into account. The integration was performed with an increased
density of points near the K-H line along which the small Fermi surface is stretched,
in order to improve the energy resolution at low frequencies.
The four tight-binding bands of graphite in the vicinity of the
K-H line are presented in figure \ref{FigBands}.
Near the H point, the bands are conical, as in monolayer graphene, while
they acquire a small mass, of the order of 5 percent of the
free electron mass, as in double-layer graphene as one moves towards the K point.
The two bands depicted in black and blue disperse strongly along the
c-axis (by about 1.5 eV), while for the other two bands (red
and green) the dispersion is only 40 meV. This latter
dispersion is responsible for the fact that the Fermi surface
is electron like at the K point and hole like at the H point.
Overall, this picture is in agreement with the recent ARPES
\cite{ZhouNP06,GruneisCM07} and de Haas-van Alphen
\cite{LukyanchukPRL04} measurements.

The in-plane conductance of graphite computed for this band
structure is shown in figure \ref{FigG1}b. One can see that the
simple band calculation is sufficient to understand the
survival of the universal conductance value in the mid-infrared
range. There are two important factors that favor this. First,
the energy of the Dirac point varies only weakly along the K-H
line. We note that in doped graphene ($\mu\neq 0$) the
conductance is gapped below $\omega = 2|\mu|$
\cite{GusyninPRL06}. The position of the Dirac energy with
respect to the chemical potential changes from about -20 meV at
the H point to about 20 meV at the K point, which explains why
the interband conductance precipitates dramatically below 40
meV (at $T$=0) and forms a peak-like structure marked by 1 and
2. Importantly, this energy scale strongly depends on the
type of stacking. For example, a similar calculation on a hypothetical
AA-stacked graphene (not shown here) indicates that in this
system the conductance would be strongly suppressed below 1.5 eV.

The second factor is a less obvious but an equally essential one.
The frequency-independent conductance can be easily explained only in
the case of non-split conical bands, while in graphite the bands are split and slightly
parabolic, except close to the H point.
In fact, a partial conductance (not shown) calculated for all momentum states with a certain
value of $k_{z}$ away of the H point has a strong frequency dependence, very similar to the expected 
conductance of bilayer graphene \cite{NilssonPRL06,AbergelPRB07}. It shows a sharp double peak due to the
transitions marked as 3 and 4 and a depletion at lower energies where only two
bands out of four contribute to it. Essentially, the optical weight
is redistributed by the c-axis hopping. However, in the total (i.e. the $k_{z}$ integrated) conductance
this redistribution is almost averaged out, due to the fact that the splitting size changes continuously from
the maximum value (0.7 eV for the transition 3 and 0.9 eV for the transition 4) at the K
point to zero at the H point. Only close to the K point, where the bands show a van Hove singularity, these
transitions make two broad conductance peaks 3 and 4
\cite{TaftPR65,GuizzettiPRL73}.

\begin{figure}[htb]
   \centerline{\includegraphics[width=4.5cm,clip=true]{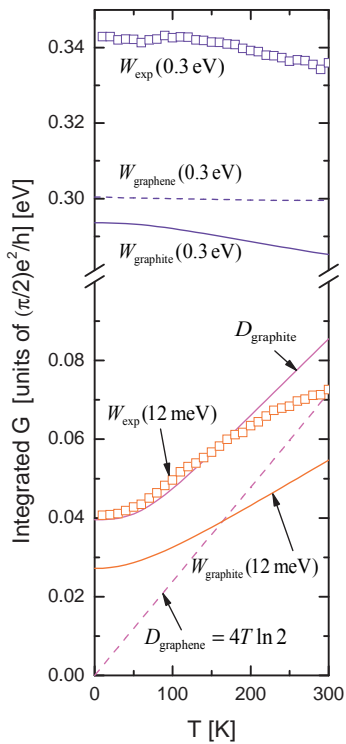}}
   \caption{The integrated conductance of graphite (symbols - experiment, solid lines -
   calculation) and of monolayer undoped graphene (calculation, dashed
   lines).}
   \label{FigSWT}
\end{figure}

Finally, we study the redistribution of the optical spectral weight between different energy
regions using the f-sum rule analysis. The red and the blue
symbols in figure \ref{FigSWT} show the measured temperature dependence of the low-energy integrated conductance
$W(\omega_{c})=\int_{0}^{\omega_{c}}G_{1}(\omega)d\omega$ for $\omega_{c}$ = 12 meV and 0.3 eV respectively.
The first one consists mostly of the Drude
(intraband) spectral weight which strongly increases as a
function of temperature due to the thermally excited
electron-hole pairs. The second one contains both the Drude
weight and the low-energy transitions between the hole and
electron bands. The fact that it is practically independent of
temperature proves that the Drude spectral weight is at the
expense of the suppression of the low-lying interband
conductance. The observed temperature dependence agrees nicely
with the described above calculations in graphite (solid
lines). The absolute experimental values are somewhat higher
than the theory predicts, which might be due to some
underestimation of the Drude weight by the SWMcC model with the
particular values of the hopping parameters used as well as due
to the absolute experimental uncertainty of $G_{1}(\omega)$.
For the purpose of illustration, we show on the same graph the
calculated pure Drude spectral weight $D(T)$ of graphene (described by
equation \ref{Dgraphene}) and of graphite, although it would
not be trivial to extract this exact value from the
experimental data, due to an overlap between the Drude peak and
the interband transitions. One can see, however, that the
calculated temperature dependence of $W(12 \mbox{ meV})$ in
graphite reflects well the one of the Drude weight alone.

In conclusion, our study reveals remarkable similarities
between the measured optical conductance of highly ordered
pyrolytic graphite per graphene layer and theoretical predictions for the monolayer
graphene. First, the optical conductance of graphite due to the transitions
between hole and electron bands is very close to the universal value of $e^2/4\hbar$ between 0.1 and 0.6 eV.
Second, the optical spectral weight removed from the low-energy interband conductance at finite temperatures
is transferred to the Drude peak. This implies that the low-energy charge dynamics in graphite is rather non-trivial
as it involves simultaneously the Drude component and the transitions between the hole and electron bands.
This work was supported by the Swiss National Science Foundation through the National Center of Competence in
Research "Materials with Novel Electronic Properties-MaNEP". We
thank N.P. Armitage, D.N. Basov and Y. Kopelevich for useful discussions.


\begin{references}
\bibitem{RemarkCalculation}
The calculations for monolayer graphene, shown in Fig.\ref{FigG1}c and \ref{FigSWT}, take the finite bandwidth in account. 
However, in the shown frequency range, $G_{1}(\omega,T)$ almost coincides with 
$G_{0}\left[f\left(-\frac{\hbar\omega}{2}\right)-f\left(\frac{\hbar\omega}{2}\right)\right]$, which is the result for perfectly conical bands.
\end{references}
\end{document}